# Demonstration of Angle Dependent Casimir Force Between Corrugations

A.A. Banishev[1], J. Wagner[1], T. Emig[2], R. Zandi[1], and U. Mohideen[1]

[1]Department of Physics and Astronomy, University of California, Riverside, California 92521, USA

[2]Laboratoire de Physique Théorique et Modéles Statistiques, CNRS UMR 8626, Université Paris-Sud, 91405 Orsay, France

Abstract

The normal Casimir force between a sinusoidally corrugated gold coated plate and a sphere was measured at various angles between the corrugations using an atomic force microscope. A strong dependence on the orientation angle of the corrugation is found. The measured forces were found to deviate from the proximity force approximation and are in agreement with the theory based on the gradient expansion including correlation effects of geometry and material properties. We analyze the role of temperature. The obtained results open new opportunities for control of the Casimir effect in micromechanical systems.

The Casimir effect [1] has seen rapid theoretical and experimental progress [2-6] in the last decade due both to its fundamental and technological applications. In its archetypal form it is an attractive force between two neutral conducting surfaces placed in vacuum. Despite its quantum nature, associated with the scattering of zero-point photons, the Casimir force between neutral metallic surfaces is a macroscopic phenomenon and can be observed at the micron scale. The Casimir force between two objects can be viewed as the collective interactions of the charge and current fluctuations induced by vacuum photons on the two bodies. Thus altering the length scale of these fluctuations using object geometry, dielectric properties and temperature lead to profoundly interesting effects. Corrugated boundaries are of interest due to the diffraction type coherent scattering effects of zero point photons that have been reported in these systems [7-12]. The photon wavelengths of interest are those that correspond to the separation distance. Other characteristic length scales in the problem are the corrugation period $\Lambda$, separation between the corrugations $z$, the thermal wavelength $\lambda_T=\hbar c/k_B T$, and the material reflectivity through the plasma wavelength $\lambda_p=2\pi c/\omega_p$. The interplay of the different length scales and the angle between the two corrugations lead to a rich behavior in this system, making it a promising probe of these coupled phenomena [13-18].

The normal Casimir force acting in the direction perpendicular to the interacting surfaces has been measured for sphere-plate configurations [19-30]. Periodically corrugated surfaces allow to better understand the macroscopic geometry effects of vacuum fluctuations. The geometry effects are usually discussed in terms of deviation from the proximity force approximation (PFA). In the simple PFA, curved surfaces are treated as a collection of flat infinitesimal surface elements and the Casimir energy is estimated as an additive sum of the local parallel plate Casimir energies. However, Casimir forces are non-additive and the PFA neglects diffraction effects and correlations from the interplay of geometry, material properties and temperature. The normal Casimir force between one corrugated surface and a sphere have been studied [8, 12, 31]. In Ref. [31] an atomic force microscope (AFM) was used to measure the normal Casimir force between a sphere and sinusoidal corrugation. In [8, 12] a microtorsional oscillator was used to measure the normal Casimir force between a sphere and rectangular corrugations and deviation from PFA demonstrated. For two aligned corrugated surfaces the lateral Casimir force [31] from the phase shift of the corrugations has been measured [9-11] and its geometry dependence demonstrated with an AFM [10, 11].

In this Letter we report the first demonstration of the angle dependence of the normal Casmir force between two corrugations. The interaction is measured between an Au coated sinusoidally corrugated sphere and plate for different orientation angles between the corrugations. We find that the normal force measurement between two sinusoidally corrugated surfaces show the interplay of thermal, material and geometry effects. Note that the corrugations used are shallow and smooth. The obtained data are compared with theoretical calculations using the derivative expansion [33] at 300 K. For the corrugation periods used, the Casimir force changes with orientation angle, increasing by 15% at 130 nm for an angle change from 0° to 2.4°. The measured forces are consistent with theory at non-zero temperature. The strong observed dependence on the corrugation angle means, that this feature can be used in adjusting and controlling moving parts in proposed micromechanical devices using corrugated surfaces and the Casimir effect [34-36].

A schematic diagram of the experimental setup is shown in Fig. 1 (please also see Ref. [37]). To perform the Casimir force measurement, two aligned corrugated surfaces, one planar and one spherical are required. The first planar surface is a diffraction grating with uniaxial sinusoidal corrugations of period $\Lambda$=(570.5±0.2) nm and amplitude $A_1$=40.2±0.3 nm. The diffraction grating is covered with a 300 nm gold coating. A 1×1 cm piece of the grating was placed on top of a rotatable stage, mounted on top of the AFM piezo. To make electrical contacts to the corrugated plate a thin copper wire is soldered to a corner using indium wire. The corrugated plate was used as a template for the in situ pressure imprinting of the corrugations on the bottom surface of a sphere. Prior to the pressure imprinting, a polystyrene sphere is attached to the tip of a 320 μm long triangular silicon nitride cantilever of nominal spring constant ~0.01 N/m using conductive Ag epoxy. Next the cantilever-sphere system was uniformly coated with a 10 nm Cr layer followed by a 20 nm Al layer and finally with a 110±1 nm Au layer using an oil-free thermal evaporator at $10^{-7}$ Torr vacuum. The radius of the Au-coated sphere was determined using a SEM to be $R$=99.6±0.5 μm. The vacuum chamber with the corrugated sphere-plate system was pumped down to a pressure below $10^{-6}$ Torr. Liquid $N_2$ cooling was used to lower the noise. Next the corrugations from the plate were pressure imprinted on the sphere to obtain two aligned corrugations. To accomplish this, the sphere was first brought into near-contact with the corrugated plate using the AFM stepper motor. Next, the top of the cantilever with attached sphere was mechanically supported using a stepper motor controlled metal stylus. Then the AFM piezo supporting the plane was extended, to make the imprint on the sphere. The amplitude of the imprinted corrugations was measured using an AFM to be $A_2$=14.6±0.3 nm and the size of imprint area was measured with an SEM to be $L_x \sim L_y \sim 14$ μm after completion of the Casimir force experiments. For changing the orientation angle between the corrugations, the corrugated plate was rotated using a stage with a third stepper motor. The stage was independently calibrated and the uncertainty in angle was determined to be 0.1°.

The corrugated plate is connected to a voltage supply operating with 1 μV resolution. The cantilever with attached sphere is grounded. With the two corrugations aligned, 11 different voltages $V_i$ in the range from -40 to -145 mV were applied to the plate. The cantilever deflection $S_{def}$ signal due to the total force (electrostatic and Casimir) between the sphere and plate was measured as a function of sphere-plate separation distance. Starting at the maximum separation of 2 μm, the plate was moved towards the sphere using a 0.05 Hz triangular voltage applied to the AFM peizo and the corresponding cantilever deflection was recorded every 0.2 nm till the plate contacted the sphere. The mean separation between sphere and plate $z$ can be written as [37]:

$$z = z_{piezo} + mS_{def} + z_0, \tag{1}$$

where $z_{piezo}$ is the movement of the plate due to the piezo (calibrated interferometrically [38]), $mS_{def}$ is the change in separation distance due to cantilever deflection, $m$ is the cantilever deflection in nm per unit photodetector signal and $z_0$ is the mean separation on contact of the two corrugated surfaces.

The cantilever deflection $S_{def}$ from the total force is given by:

$$S_{def}(z) = \frac{F_{tot}(z)}{k'} = \frac{F_{el}(z) + F_{Cas}(z)}{k'} = \frac{X(z)}{k'}(V_i - V_0)^2 + \frac{F_{Cas}(z)}{k'}. \tag{2}$$

$F_{Cas}$ and $F_{el} = X(z)(V-V_0)^2$ are the Casimir and electrostatic forces between sphere and grating. $V_0$ is the residual potential difference between the surfaces. The quantity $k' \equiv km$ is the cantilever calibration constant measured in units of force per unit deflection (pN/mV), where $k$ is the cantilever spring constant. The coefficient $X(z)$ is calculated by using PFA to account for the curvature of the sphere and by using a perturbative expansion in powers of the amplitudes $A_1$ and $A_2$ to treat the corrugations:

$$X(z) \approx \frac{\pi R}{z} + \frac{\pi R}{z^2}\left[\frac{\pi}{\Lambda}(A_1^2 + A_2^2)\coth(2\pi z/\Lambda) - \frac{4\pi A_1 A_2}{\Lambda}\frac{\exp(-2\pi z/\Lambda)}{1-\exp(-4\pi z/\Lambda)}\frac{\sin(\pi L_y \theta/\Lambda)}{\pi L_y \theta/\Lambda}\right] \tag{3}$$

where $R$ is the radius of the sphere, $\Lambda$ is the period of the corrugations, $\theta$ is the crossing angle between the corrugations, and $L_y$ is the extent of the corrugations along the axis of the corrugations. Eq. (3) was checked with a numerical computation of the electrostatic force using a finite element method for separations between 160 to 400 nm, and was shown to agree to better than 1%.

After measuring the deflection $S_{def}$ due to the total force, we subtracted any mechanical drift of the photodetector system with respect to the cantilever, determined the point of sphere–plate contact and the cantilever deflection coefficient $m=102.1\pm0.5$ nm/unit deflection signal as described in Ref. [37, 39]. The value of $m$ was used to calculate the change in separation $mS_{def}$ due to the cantilever deflection in Eq. (1). $V_0$, $k'$, and $z_0$ are found from the parabolic dependence of the electrostatic force on the applied voltage as described in Ref. [37]. The cantilever deflection was determined at intervals of 1 nm using linear interpolation. At each sphere-plate separation, the deflection $S_{def}$ was plotted as a function of the applied voltage. The vertex of the generated parabolas corresponds to $V_0$ and is determined by least $\chi^2$ fitting. The curvature of the parabolas at every separation corresponding to $X(z)/k'$ is fit using Eq. (3) to determine $z_0$ and $k'$. The measurements were repeated 10 times leading to 110 forces at each separation. For $\theta=0°$ between the corrugations the obtained values were found to be $V_0=-(90.2\pm1.3)$ mV, $z_0=(126.2\pm0.4)$ nm, and $k'=(1.35\pm0.02)$ pN/mV. The values of $V_0$, $z_0$, and $k'$ are found for each orientation angle and were confirmed to be independent of separation. From the value of $z_0$ the absolute separation distance can be determined and the value of $k'$ was used to convert $S_{def}$ to a force.

Using the calibration parameters above, the Casimir forces are obtained by subtraction of the electrostatic force from the total measured force as: $F_{cas}(z)=k'S_{def}(z)-X(z)(V_i-V_0)^2$. The mean value of the normal Casimir force is shown as crosses in Fig. 2. The size of the cross corresponds to the total horizontal and vertical errors at 67% confidence level. The details of the experimental error analysis procedure can be found in Ref. [24, 37, 40]. For $\theta=0°$, the random error was 0.51 pN and separation independent. The systematic error in the Casimir force ranged from 0.79 to 0.64 pN for separations from 127 to 300 nm. The total error in the Casimir force was found to range from 0.94 to 0.82 pN for separations from 127 to 300 nm. The Casimir force at a distance of 130 nm increases in the order 84.9, 88.8, 92.5 and 97.8 pN for orientation angles of 0, 1.2, 1.8 and 2.4° respectively for a total change of 15%. Note that this angle dependence is a finite size effect. For larger angles, the multiple crossings of the corrugations will lead to negligible angle dependence.

To compare the experiment with the theoretical Casimir force between the corrugated sphere and plate, we need to take into account the geometric features of the sphere-plate configuration as well as the corrugations. Different approximations are used for the two features. The sphere-plate geometry is treated using the PFA as $F_{Cas}=2\pi R U_{corr}$, where $R$ is the sphere radius and $U_{corr}$ is the Casimir energy per unit area for two parallel corrugated plates at angle $\theta$. The PFA is valid when the radius of curvature is much larger than the separation. The effect of corrugations on the Casimir force appear in the $U_{corr}$ term. The Casimir energy per unit area between corrugated plates can be calculated using the derivative expansion since the surfaces are gently curved because $\Lambda$ is large compared to $A_1$, $A_2$ [33, 41, 42]. For an area $L_x L_y$ the Casimir energy divided by this area is then given by:

$$U_{corr} \approx U_{PFA} + \frac{1}{L_x L_y} \int_{-L_x/2}^{L_x/2} dx \int_{-L_y/2}^{L_y/2} dy (\alpha(H)\nabla H \cdot \nabla H - \frac{1}{2}(HU'(H)-U(H))\nabla h_1 \cdot \nabla h_2)$$

$$U_{PFA} = \frac{1}{L_x L_y} \int_{-L_x/2}^{L_x/2} dx \int_{-L_y/2}^{L_y/2} dy U(H) \qquad (4)$$

where $H(z,x,x')=z+h_1(x)-h_2(x')$ is the local separation distance between the plates with $h_1(x)=A_1 cos(2\pi x/\Lambda)$, $h_2(x')=A_2 cos(2\pi x'/\Lambda)$ the profiles of the two plates and $x'\equiv x cos\theta - y sin\theta$. Note that for infinite plates this energy has no angular dependence on $\theta$. Hence we consider angles $\theta$ for which the periodicity $\Lambda/sin\theta$ is larger than or comparable to $L_y$. The first term $U_{PFA}$ in Eq. (4) gives the PFA for the corrugation, and the remaining terms are the corrections in a derivative expansion. The functions $U(d)$ and $U'(d)$ are the expression for the Casimir energy per unit area between two parallel plates separated by a distance $d$ and its derivative. These expressions are given by the usual Lifshitz formula and its partial derivative with respect to separation, respectively. The function $\alpha(d)$ is the coefficient of the first correction to the PFA and can be calculated as in Ref. [33]. The material properties are introduced in the calculation of $U(d)$, $U'(d)$, and $\alpha(d)$ in perturbative height expansion of the reflection coefficient [33].

The theoretical computation of the Casmir forces was performed with the real properties of Au at 300 K. The permittivity of Au was expressed using the 6-oscillator model for the core electrons and the Drude model for the free electrons, in terms of the imaginary frequency $\zeta$ [4] as:

$$\varepsilon(i\zeta) = 1 + \frac{\omega_p^2}{\zeta(\zeta+\gamma)} + \sum_{i=1}^{6} \frac{f_i}{\omega_p^2 + \zeta^2 + \zeta g_i}. \tag{5}$$

The plasma frequency $\hbar\omega_p$=9 eV, the relaxation constant $\hbar\gamma$=0.035 eV, and the oscillator constants found in Ref. [4] were used. For the experimental errors here, the difference between $\gamma\neq 0$ (Drude model) and $\gamma=0$ (plasma model) cannot be discerned. The small roughness corrections were done as described in Ref. [11]. The roughness over 20 corrugation periods was found by analyzing the difference between the measured AFM profiles and an ideal sine curve. The rms roughness was found to be $\delta_1$=1.9 nm and $\delta_2$=2.9 nm for the corrugated plate and sphere respectively.

The comparison of the experimental data with theory is shown in Fig. 2. Good agreement is found. No fitting parameters are used. In the inset of Fig. 2, for $\theta$=0, the deviation from PFA is explored by plotting the ratio of the experimental data to the force obtained from PFA (corresponding to $U_{PFA}$). The largest deviation is 7.7%. Note that such a deviation is observed even for the shallow, smooth corrugations used. To further illustrate the role of correlation effects and the interplay of the material properties on the geometry of the periodic corrugations, the experimental data was compared with the difference force obtained by subtracting the theoretical PFA force from the measured values. This is displayed in Fig. 3a and 3b for angles 0° and 1.2°  between the corrugations. Here the difference between the experimental data and the PFA is compared to the difference force between the derivative expansion and the PFA, corresponding to $U_{corr}$ - $U_{PFA}$. The error bars (at 67% confidence level) represent the data and the theoretical difference is represented by the solid line. One can observe that there is a significant deviation of the experimental data from the theory based on simple PFA. For the separation of 130 nm the absolute deviation is 5.9 pN and 4.2 pN for angles $\theta$=0° and 1.2°, respectively, which is more than twice the error bar. As can be observed in Fig. 3 the solid lines representing the deviation of the derivative expansion theory from PFA are in good agreement with the deviation from PFA in the experiment. The role of the material dependence is explored by additionally comparing the theoretical difference force for an ideal metal at 300 K as the dotted curves in Fig. 3. The apparent agreement in Fig. 3(b) is a numerical coincidence and is part of a trend, where the theoretical force difference for an ideal metal increases with angle (the difference is more telling for the other angles as shown in the supplementary materials). Note that the total Casimir force is always larger for the perfect metal. To explore the role of temperature the ratio of the experimental data to the force from the derivative expansion at 300 K and 0 K is shown in the inset. The data is found to be consistent with 300 K.

In conclusion, we have experimentally demonstrated the angle dependence of the normal Casmir force between a corrugated plate and corrugated sphere. The measured Casimir force was shown to increase by 15% at 130 nm separation when the orientation angle between corrugations is increased from 0° to 2.4°. The measurements were found to be in agreement with theory based on the derivative expansion and shown to deviate significantly when the corrugations are treated only with the simple PFA. The experimental agreement when deviations from PFA along with real material properties are included, demonstrate the interplay of the correlation effects of the geometry with the material properties. The angle dependent Casimir force for two oriented corrugations is an important system for understanding the non-trivial combined interactions of geometry, material properties and temperature. This demonstration for corrugated surfaces will find applications in adjusting and controlling the functionality of closely spaced moving parts of micromachines.

The authors acknowledge extensive discussions with M. Kardar. This work was supported by NSF Grant No. PHY 0970161 (U.M, A.B).

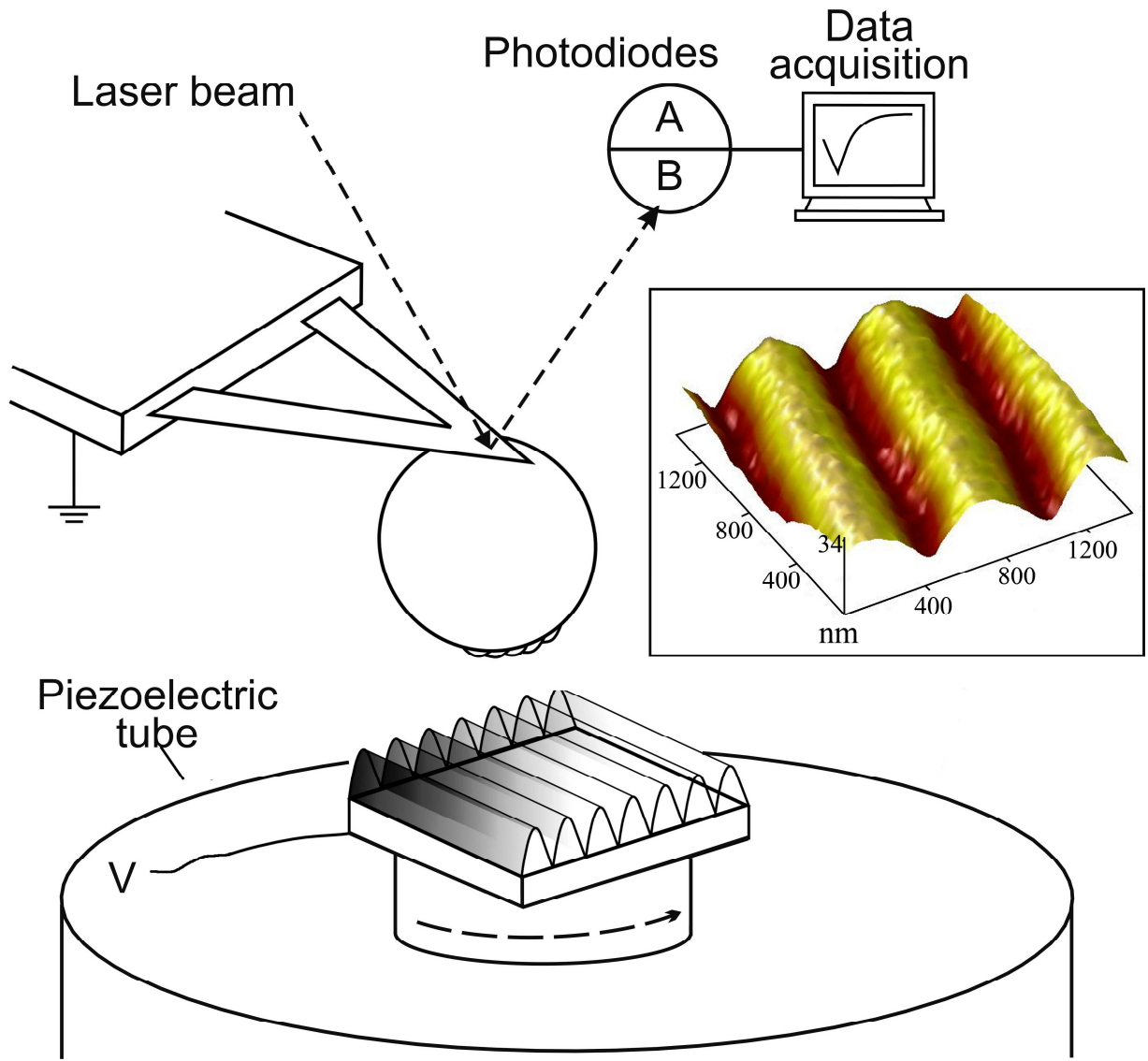

Fig. 1: Schematic of the experiment setup. Inset is the AFM image of the imprinted corrugations on the sphere.

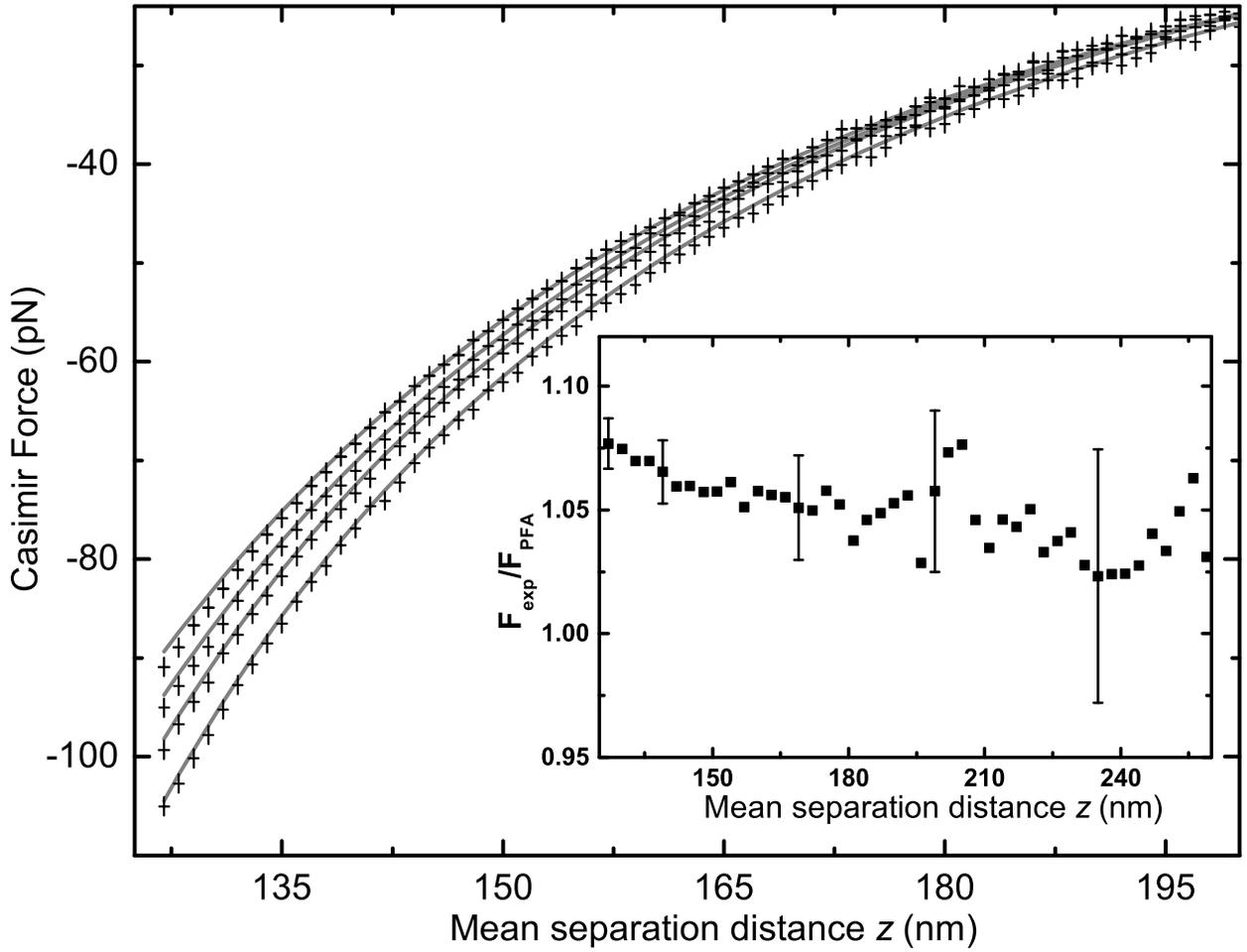

Fig. 2: Measured Casimir forces for different orientations of the corrugations. Data shown as crosses, from the top to the bottom correspond to angles of $0^o$, $1.2^o$, $1.8^o$ and $2.4^o$ respectively. The size of the crosses represents the total error. Solid lines represent the derivative expansion theory. Inset is the ratio of the data to the PFA force at $\theta=0°$.

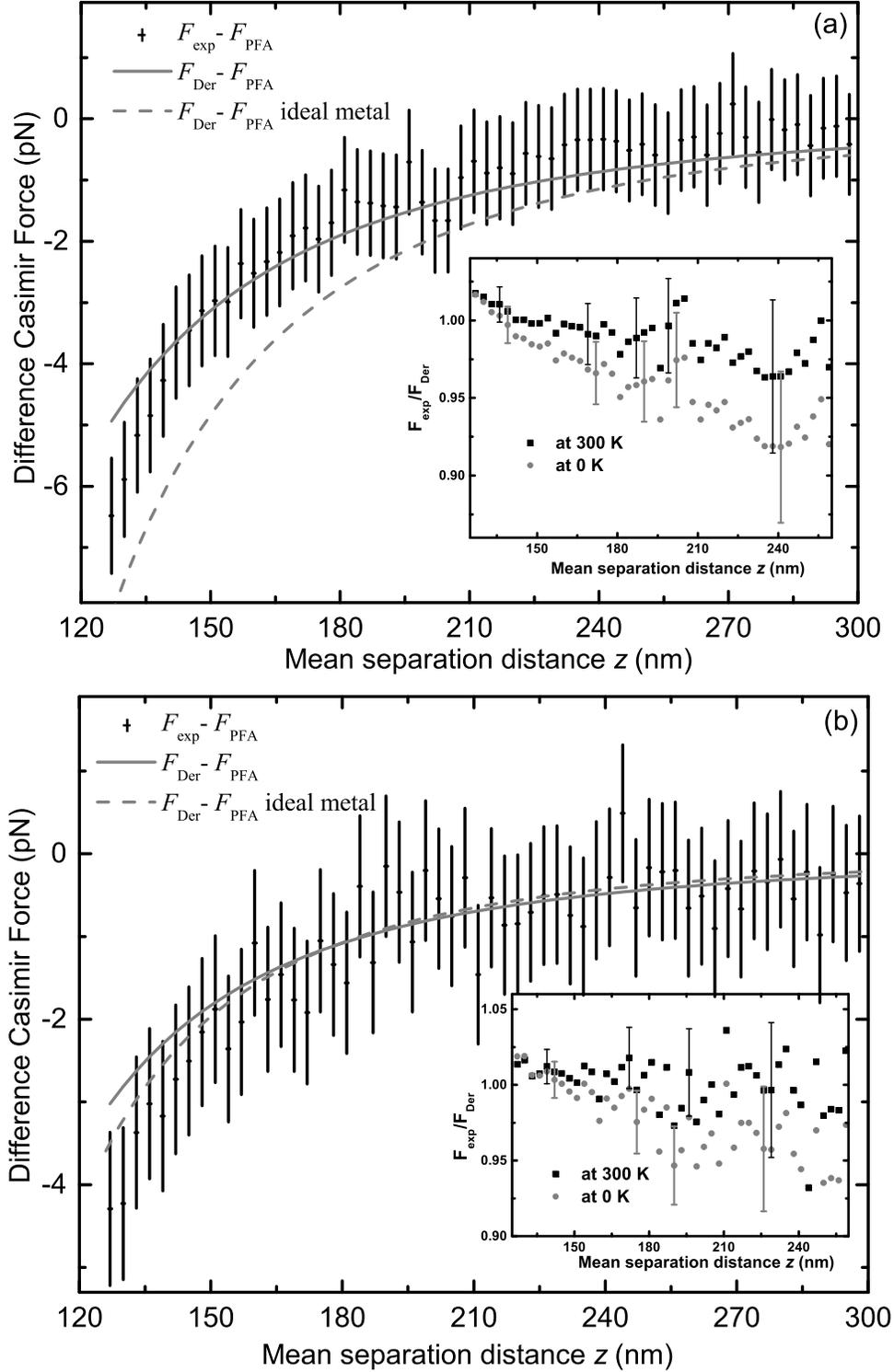

Fig. 3: Difference Casimir force $F_{exp}$ - $F_{PFA}$ represented as crosses corresponding to error bars for $\theta$ (a) 0° and (b) 1.2°. Solid line is the difference between the two theories $F_{Der}$–$F_{PFA}$, which is a measure of correlation effects. Dashed line is the theoretical difference for ideal metal corrugated

surfaces at 300 K. The data is presented every 3 nm for clarity. Inset shows the ratio of the data to the force from the derivative expansion at 300 K (black squares) and 0 K (grey circles).

## Supplementary Materials

To further elucidate the role of correlation effects and the interplay of the material properties on the geometry of the periodic corrugations, the experimental data are compared with that corresponding to the results obtained using a perfect metal in the derivative expansion for the other two angles $1.8°$ and $2.4°$ not shown in the main body of the paper. The corresponding force difference is shown below in Fig. 1 (a) and (b) respectively. The difference between the experimental data and the PFA estimate is compared to the difference force between the derivative expansion and the PFA, corresponding to $U_{corr} - U_{PFA}$. The error bars (at 67% confidence level) represent the data and the theoretical difference is represented by the solid line. The dashed line is the difference force where the corrugations are assumed to be made of a perfect metal. The large deviation from the data for both angles from that due to ideal metals shows the role of the material dependence. Please note that the force difference is plotted and the total Casimir force is always larger for ideal metals.

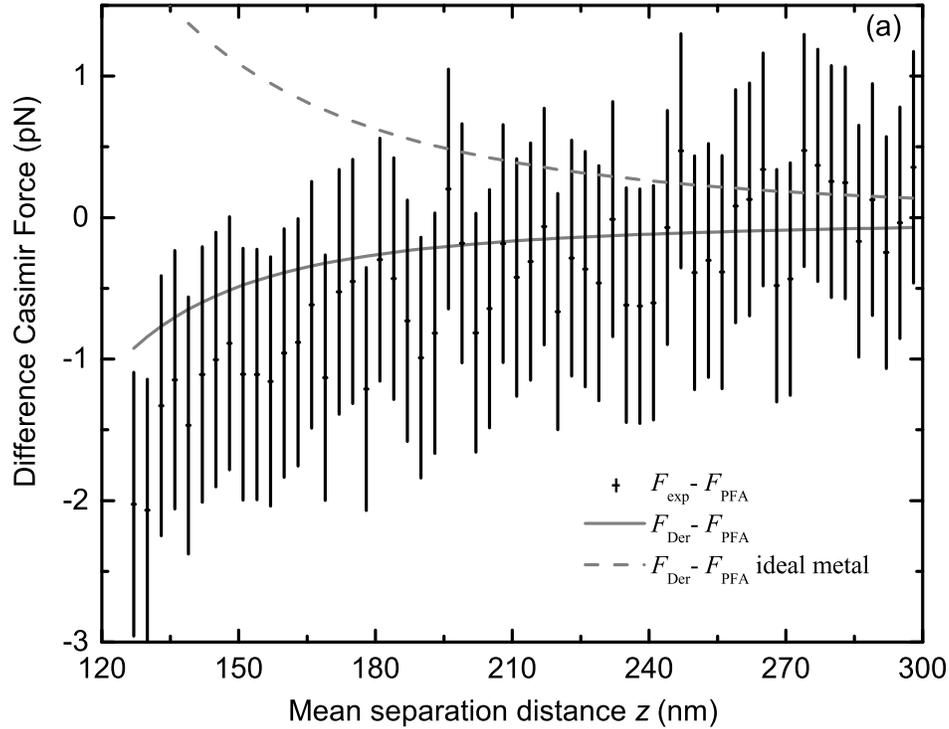

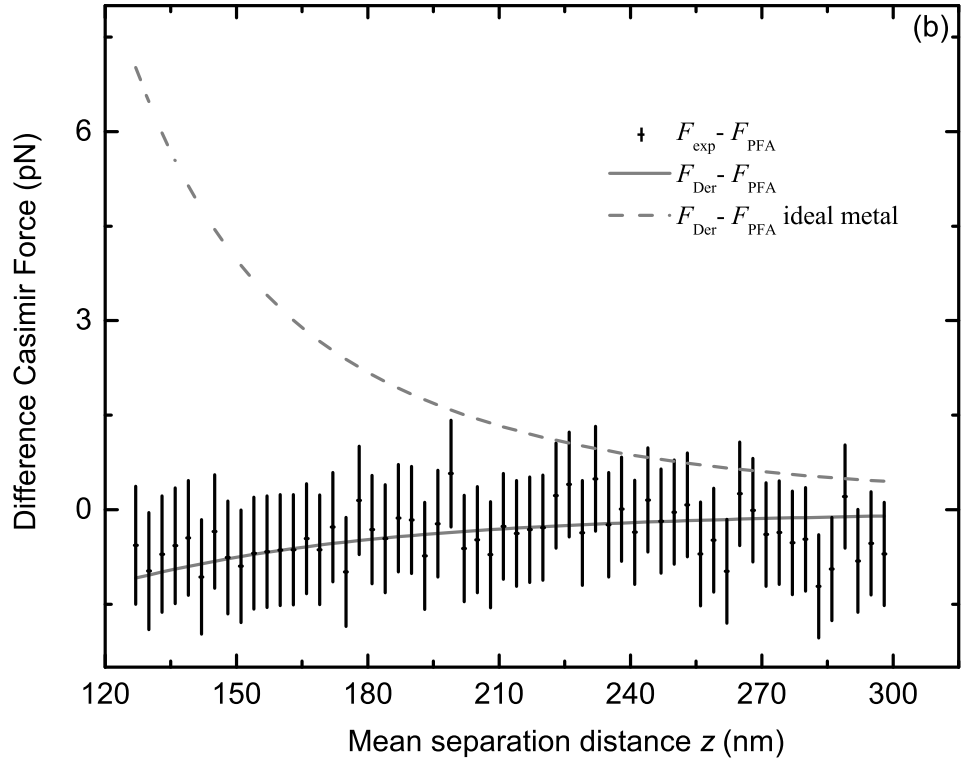

Figure 1: The difference Casimir force $F_{exp} - F_{PFA}$ represented as crosses corresponding to error bars at 67% confidence level for corrugation orientation angles of (a) 1.8° and (b) 2.4°. The solid line is the corresponding difference between the two theories $F_{Der} - F_{PFA}$, which is a measure of the correlation effects. Deviation from the simple PFA is observed and the good agreement with the complete theory is found. The dashed line is the theoretical difference for ideal metal corrugated surfaces at 300 K. The data is presented every 3 nm for clarity.